\documentclass[lettersize,journal]{IEEEtran}
\usepackage{balance}
\usepackage{amssymb}
\usepackage{stmaryrd}
\usepackage{color}
\usepackage{multirow}
\usepackage{graphicx,times}
\usepackage{epstopdf}
\usepackage{indentfirst}
\usepackage{CJK}
\usepackage{amsmath}
\usepackage{amsfonts}
\usepackage{txfonts}
\usepackage{mathrsfs}
\usepackage{subfigure}
\usepackage{graphicx}
\usepackage{theorem}
\usepackage{url}
\usepackage{microtype}
\usepackage{fancyhdr}  
\usepackage{booktabs} 
\usepackage{booktabs}   
\usepackage{tabularx}  
\usepackage{array}
\usepackage{float}

\begin{document}
	
\title{Securing UAV Communications by Fusing \\ Cross-Layer Fingerprints}

\author{Yong Huang,~\IEEEmembership{Member,~IEEE}, Ruihao Li, Mingyang Chen, Feiyang Zhao, Dalong Zhang, and \\ Wanqing Tu,~\IEEEmembership{Senior Member,~IEEE}
\thanks{This work was supported in part by the National Natural Science Foundation of China with Grant 62301499 and the Henan Association for Science and Technology with Grant 2025HYTP037.
Wanqing Tu’s work is supported by the EPSRC and DSIT through the Communications Hub CHEDDAR (grant numbers EP/X040518/1 and EP/Y037421/1), and European Union’s Horizon Europe research and innovation programme grant “SENSORBEES” (agreement no. 101130325).
\textit{(Corresponding author: Wanqing Tu.)}}
\thanks{Y. Huang, R. Li, M. Chen, F. Zhao, and D. Zhang are with the School of Cyber Science and Engineering, Zhengzhou University, Zhengzhou 450001, China (e-mail: yonghuang@zzu.edu.cn; ruihaoli@gs.zzu.edu.cn; chenmingyangzzu@gmail.com; zhaofeiyang000@gs.zzu.edu.cn; iedlzhang@zzu.edu.cn).}
\thanks{W. Tu is with the Department of Computer Science, Durham University, United Kingdom (e-mail: wanqing.tu@durham.ac.uk).}
\thanks{Copyright (c) 2025 IEEE. Personal use of this material is permitted. However, permission to use this material for any other purposes must be obtained from the IEEE by sending a request to pubs-permissions@ieee.org.}
}

\markboth{IEEE Internet of Things Journal, ~Vol.~X, No.~X, XXXXX~XXXX}%
{Shell \MakeLowercase{\textit{et al.}}: A Sample Article Using IEEEtran.cls for IEEE Journals}

\IEEEpubid{0000--0000/00\$00.00~\copyright~2025 IEEE}

\maketitle

\begin{abstract}
The open nature of wireless communications renders unmanned aerial vehicle (UAV) communications vulnerable to impersonation attacks, under which malicious UAVs can impersonate authorized ones with stolen digital certificates. 
Traditional fingerprint-based UAV authentication approaches rely on a single modality of sensory data gathered from a single layer of the network model, resulting in unreliable authentication experiences, particularly when UAVs are mobile and in an open-world environment. 
To transcend these limitations, this paper proposes SecureLink, a UAV authentication system that is among the first to employ cross-layer information for enhancing the efficiency and reliability of UAV authentication.
Instead of using single modalities, SecureLink fuses physical-layer radio frequency (RF) fingerprints and application-layer micro-electromechanical system (MEMS) fingerprints into reliable UAV identifiers via multimodal fusion. 
SecureLink first aligns fingerprints from channel state information measurements and telemetry data, such as feedback readings of onboard accelerometers, gyroscopes, and barometers.
Then, an attention-based neural network is devised for in-depth feature fusion.
Next, the fused features are trained by a multi-similarity loss and fed into a one-class support vector machine for open-world authentication.
We extensively implement our SecureLink using three different types of UAVs and evaluate it in different environments.
With only six additional data frames, SecureLink achieves a closed-world accuracy of 98.61\% and an open-world accuracy of 97.54\% with two impersonating UAVs, outperforming the existing approaches in authentication robustness and communication overheads.
Finally, our datasets collected from these experiments are available on GitHub: https://github.com/PhyGroup/SecureLink.
\end{abstract}

\begin{IEEEkeywords}
UAV authentication, impersonation attack, UAV data link
\end{IEEEkeywords}

\section{Introduction}

The past decade has witnessed wide applications of unmanned aerial vehicles (UAVs), also known as drones, in many intriguing sectors, such as package delivery, aerial photography, light shows, and smart agriculture~\cite{r1, wei2022uav}.
Typically, a UAV system consists of UAVs, ground control stations (GCSs), and wireless links between them.
The open nature of wireless links makes them the primary source of security vulnerabilities in the UAV system~\cite{zhi2020security}.
With the increasingly crowded and hostile airspace, identity authentication is crucial to establishing reliable communications and determining which UAVs can access UAV networks or other valuable resources.
Many commercial-off-the-shelf UAVs adopt cryptography-based digital certificates as unique identifiers~\cite{tsao2022survey}.
However, due to limited airborne computational capability and lack of physical isolation, such protection is vulnerable to UAV impersonation attacks, where illegitimate UAVs pretend to be authorized with stolen digital certificates~\cite{zhi2020security}.
For instance, impersonating UAVs can disturb the coordination of aerial swarms.
Besides cryptographic schemes, a lightweight and reliable supplementary authentication approach is necessary for trustworthy aerial or aerial-ground applications.

\IEEEpubidadjcol

Visual~\cite{opromolla2018vision}, radar~\cite{r10}, or acoustic~\cite{r8} characteristics are often exploited to detect or classify flying UAVs.
However, these methods are yet unsuitable for verifying their digital IDs in wireless communication links.
This is because finding the associations between UAV communication links and their visual, radar, or acoustic profiles remains challenging, especially when multiple UAVs are present or non-line-of-sight communication happens~\cite{r9}.
In contrast, radio frequency (RF) and micro-electromechanical system (MEMS) fingerprints are promising enablers to meet this requirement.
These fingerprints stem from airborne RF chains and MEMS sensors and are unique to each UAV.
More importantly, such information is naturally contained in the physical layer and the application layer of UAV communications, making it readily accessible.
In the literature, RF signals are commonly sampled using in-phase and quadrature (IQ) techniques to fingerprint different UAVs.
However, these approaches require bulky and expensive software-defined radio equipment to be deployed on the ground ~\cite{r5,9010185,droneauth1, r7}.
Increasingly, attempts have been devoted to exploring RF fingerprinting based on channel state information (CSI) obtainable from commercial wireless network interface cards (NICs)~\cite{29,30,picoscenes}.
The literature~\cite{Micro-CSI,CSI-Phase-error,fra-CFO,23} has developed CSI-based fingerprinting for static wireless devices by extracting features, such as carrier frequency offsets and phase errors.
Also, the researchers have shown the feasibility of UAV identification using MEMS imperfection, but this is verified on stand-alone gyroscope sensors in a static condition~\cite{gro-bias}.
In addition, most of the above approaches assume a closed-world communication environment, and may not work well in an open-world setting, where unknown UAVs can appear during the authentication process.
In summary, the existing RF or MEMS fingerprint-based studies are single-layer authentication technologies. 
They mainly focus on static wireless devices and could suffer from dynamic UAV motion states in an open-world environment.

To overcome these limitations, we advocate the fusion of inherent RF and MEMS imperfections to secure UAV communications, rather than relying on a single modality of sensory data. 
This is because RF and MEMS fingerprints are complementary in terms of UAV authentication.
By combining the two types of fingerprints belonging to the physical layer and the application layer, respectively, the UAV identifiers become more reliable in overcoming the challenges caused by single-modality data, such as motion dynamics, electromagnetic interference, and measurement outliers,
improving the robustness of UAV authentication.
Moreover, because each UAV data packet contains both CSI and MEMS data, the employment of two modalities of data can increase the fingerprint-related information and reduce the amount of packets required for UAV authentication, thus decreasing communication overheads.
To the best of our knowledge, no other studies have proposed such cross-layer fingerprints to authenticate flying UAVs.

Towards this end, we present SecureLink, an efficient and reliable UAV authentication system that fuses cross-layer information in UAV wireless links.
The key idea of SecureLink is to integrate measurements of airborne RF and MEMS imperfections from Wi-Fi CSI measurements and telemetry data via multimodal fusion.
Specifically, CSI is a physical layer indicator in UAV communications and describes the propagation of wireless signals from a transmitter to a receiver, but also carries hardware impairment information of RF chains.
Telemetry data is encapsulated in the application layer~\cite{r14} of UAV data communications, which consists of sensor outputs of onboard accelerometers, gyroscopes, barometers, and so on, and is transmitted from UAVs to GCSs for status monitoring.
Taking full advantage of the two freely available sensory readings, SecureLink can facilitate UAV identification without hardware modification on the GCS.
It can extract raw RF and MEMS fingerprints, respectively, and fuse their common and exclusive information into high-dimensional UAV identifiers using a multimodal learning model.

To realize this idea, we address the following challenges.

\textit{1) How to align heterogeneous CSI and telemetry readings?} 
Due to the high sensitivity to environmental dynamics, CSI measurements have variable time intervals and are much sparser than telemetry readings, significantly hampering their periodicity and correspondence with telemetry data.
To handle this issue, we first extract useful phase errors and MEMS sensor readings from raw CSI and telemetry data.
Then, we remove outliers from the two modalities and align CSI fingerprints to telemetry fingerprints using low-complexity linear interpolation. 

\textit{2) How to perform effective multimodal fusion under UAV mobility?}
CSI and telemetry measurements are highly impacted by environment and motion changes, which, however, frequently occur in flying UAVs due to their high mobility.
To address this challenge, we design a two-branch deep neural network (DNN) that exploits a 1D convolutional neural network (CNN) with a bi-directional long short-term memory (BiLSTM) to extract unimodal features from raw RF and MEMS fingerprints and fuse them into multimodal feature vectors with a multi-head attention mechanism.  

\textit{3) How to facilitate accurate UAV authentication in the open world?}
In the real world, a UAV authentication system should have the ability to recognize unseen impersonating UAVs.
For this purpose, we train a classification model using a novel multi-similarity loss function for generating unique and discriminative UAV fingerprints.
Then, a one-class support vector matching (OC-SVM) is trained to estimate the likelihood of unknown classes, thus enabling open-world UAV authentication.

\textbf{Contributions.} The main contributions of this work are summarized as follows. 
\begin{itemize}
    \item We are among the first to show that freely available CSI and telemetry measurements convey rich information about airborne hardware imperfections and can be fused to authenticate commercial-off-the-shelf UAVs in the real world.
    Previous methods rely on a single modality of sensory data from a single network layer, leading to unreliable authentication experiences.
    \item We propose a novel cross-layer UAV authentication system, SecureLink, to accurately identify authorized and impersonating UAVs with short delays and greatly controlled traffic overheads by using a customized two-branch DNN with a multi-head attention mechanism.
    SecureLink has high robustness to UAV model variations and can easily adapt to drone manufacturers, offering high compatibility across different vendors.
    \item We develop a real-world testbed for evaluation. 
    To the best of our knowledge, this is among the first to be established using three types of heterogeneous UAVs. 
    Also, we leverage this testbed to conduct extensive experiments across four different real-world open or close environments, verifying the effectiveness and robustness of SecureLink against impersonation attacks.
    \item We publish our datasets collected from our experiments on GitHub~\cite{dataset}. These datasets comprise twelve hours of multimodal data from 22 UAVs, including the measurements of Wi-Fi CSI, accelerometers, barometers, and time-of-flight (ToF) sensors.
    There are no similar datasets publicly available in the literature that were collected using heterogeneous UAV models in different real-world environments.
\end{itemize}

\textbf{Summary of Results.} 
We implement SecureLink with twenty DJI Tello drones, one DJI Phantom 4 Pro drone with the ESP32 system-on-chip (SoC), and one DJI Phantom 4 Pro drone with the ESP32 S3 SoC.
In our real-world experiments with the above heterogeneous UAVs, we collect multimodal data lasting about twelve hours in various indoor and outdoor environments on different days.
The evaluation results demonstrate that SecureLink achieves an accuracy of 98.61\% and a true negative rate (TNR) of 99.04\% in the closed world, only requiring six additional data frames.
In our open-world experimental scenarios, SecureLink achieves an accuracy of 97.54\% and a TNR of 96.95\%.
This accuracy is achieved with a runtime of approximately 15ms per authentication request.
SecureLink achieves at least 29\% higher accuracy than existing single-layer approaches.

\section{Threat Model and Preliminary Study}\label{sec:preliminaries}

\subsection{Threat Model}
We consider a common UAV system, in which a UAV operator controls and manages UAVs via a ground control station.
Both the UAVs and the GCS have wireless communication modules that utilize IEEE 802.11 protocols.
Moreover, an upper-layer protocol, e.g., MAVLink, is adopted by them to facilitate the transmissions of sensory data and control commands~\cite{mekdad2023survey}.
According to the regulations of the Federal Aviation Administration (FAA), each UAV should have a unique ID during flight.
Before establishing communications with the GCS, each UAV has to send its ID and the corresponding digital certificate.
After passing the verification process, the UAV feeds telemetry data, such as location, altitude, and other sensor outputs, back to the GSC for status monitoring.

In our system, we consider UAV impersonation attacks, which jeopardize the authenticity of the legitimate UAVs.
We assume that an attacker has obtained the software-level digital certificate of an authorized UAV.
Then, the attacker controls an unregistered UAV to claim it as the legitimate UAV, sends an authentication request to the GCS, and passes the ID check with the stolen certificate.
The security breaches of IEEE 802.11 protocols or upper-layer protocols enable such attacks in practice~\cite{mekdad2023survey, intwala2022system}.
Hereafter, the impersonating UAV can sneak into the UAV network or access valuable network resources.
Furthermore, we do not assume the type of impersonating UAVs.
They may be the same as or entirely different from legitimate platforms.

\subsection{Theoretical Explanation of RF and MEMS Fingerprints}

\textbf{RF Fingerprints}. \label{CSI-based-fingerprints}
We extract RF fingerprints from channel state information measurements available in the physical layer of UAV communications.
Specifically, CSI is a channel indicator that describes the wireless channel characteristics from a transmitter to a receiver.
Given the transmitting and receiving signals $\mathbf{x}$ and $\mathbf{y}$, the CSI $\mathbf{h}$ can be expressed by $ \mathbf{h} =\frac{\mathbf{y}}{\mathbf{x}} + \sigma$, where $\sigma$ is the environmental noise. 
Due to hardware imperfections, like IQ imbalance and oscillator offsets, the received signal $\mathbf{y}$ contains the impairment information of the transmitter's RF chain.
We extract nonlinear phase errors from CSI measurements as our RF fingerprints.
In particular, the phase error $\mathbf{e}$ can be calculated as $\mathbf{e}=\mathbf{\phi} - \mathbf{\psi} - 2\pi \lambda \mathbf{i} - \mathbf{z}.$
Therein, $\mathbf{\phi}$ denotes all subcarrier phases of the receiver.
$\mathbf{\psi}$ is the true phase of the transmitter.
$\lambda$ is a constant and represents the sum of phase offsets due to frame detection, sampling frequency offset, and time of flight.
$\mathbf{i}$ denotes the subcarrier index sequence.
$\mathbf{z}$ is another constant indicating the center frequency phase offset caused by the time of flight.
Thus, the phase error $\mathbf{e}$ contains hardware imperfections.

\textbf{MEMS Fingerprints}. Micro-electromechanical system technology enables the integration of micrometer- and nanometer-scale mechanical structures on microchips, which are widely used to fabricate various airborne sensors. 
Subtle defects in such sensor chips will inevitably occur due to imperfections in the electromechanical structure during manufacturing.
Although these defects distort the sensor outputs slightly, they are unique to each sensor and can be considered as hardware fingerprints~\cite{9951057}.
Typically, MEMS sensors employ an analog-to-digital (ADC) module to sample analog signals and store them in digital registers. 
For example, the output of a common triaxial motion sensor $\mathbf{o}$ can be represented by the following equation~\cite{acc-factory-calibration}
\begin{equation}\label{eq:sensors error}
\mathbf{o}=\begin{bmatrix}o_x\\o_y\\o_z\end{bmatrix}=
\begin{bmatrix}c_x\\c_y\\c_z\end{bmatrix}
\begin{bmatrix}1&d_{xy}&d_{xz}\\d_{yx}&1&d_{yz}\\d_{zx}&d_{zy}&1\end{bmatrix}
\begin{bmatrix}a_x+b_x\\a_y+b_y\\a_z+b_z\end{bmatrix},
\end{equation}
where $c_x$ is the scale factor on the X-axis.
$d_{xy}$ represents the non-orthogonality between the X and Y axes.
$a_x$ is the ADC output of this sensor, and $b_x$ is the bias.
The above equation indicates that the outputs of airborne MEMS sensors contain rich hardware imperfections, laying the foundation for authenticating UAVs based on MEMS fingerprints.

\begin{figure}
    \centering
    \subfigure[Phase errors in the stationary state.]{
    	\includegraphics[width=0.46\columnwidth]{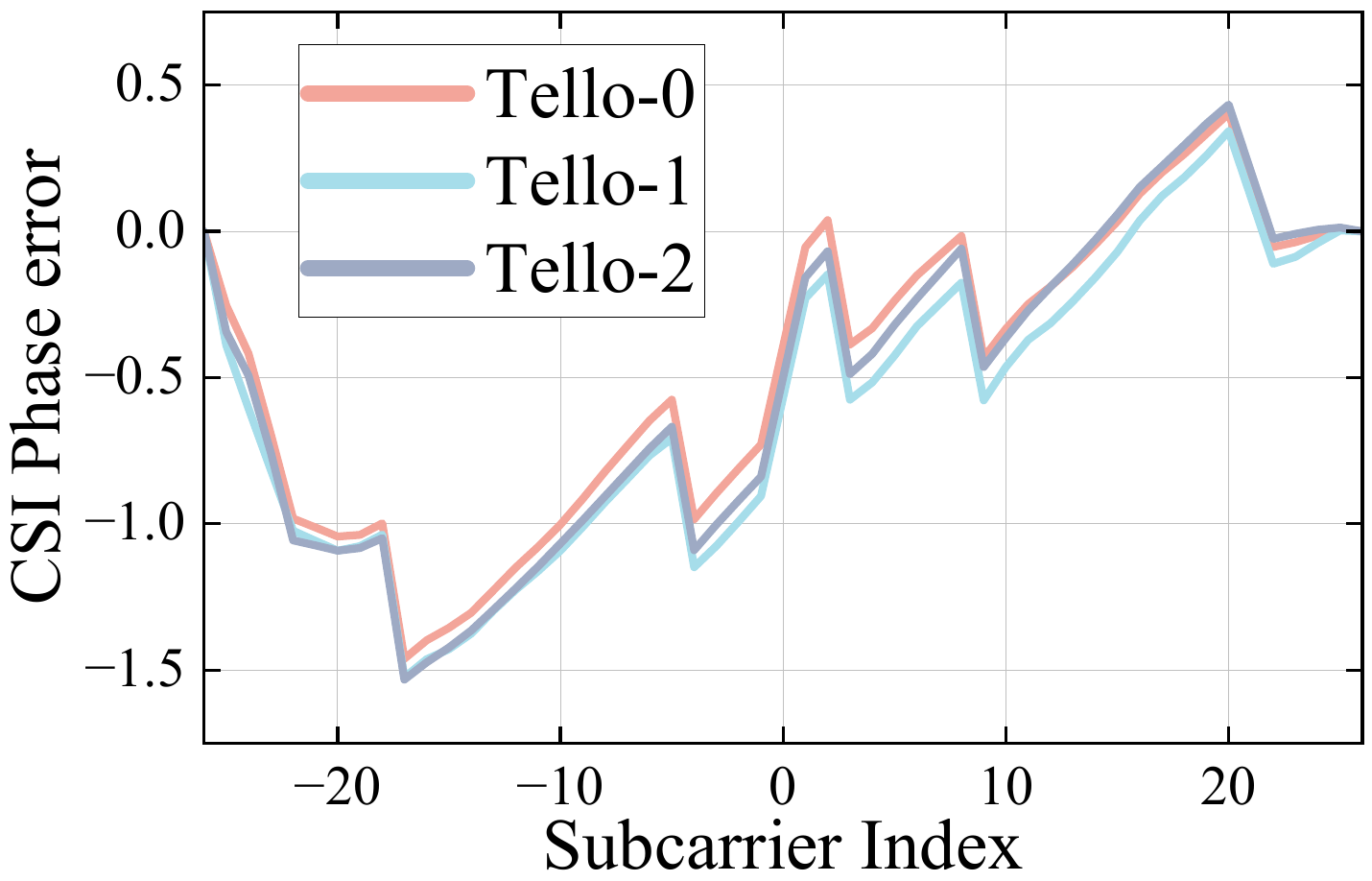}
    	\label{fig:Tello_CSI_phase_errors_static}
     }
     \subfigure[Phase errors in the flying state.]{
     \includegraphics[width=0.46\columnwidth]{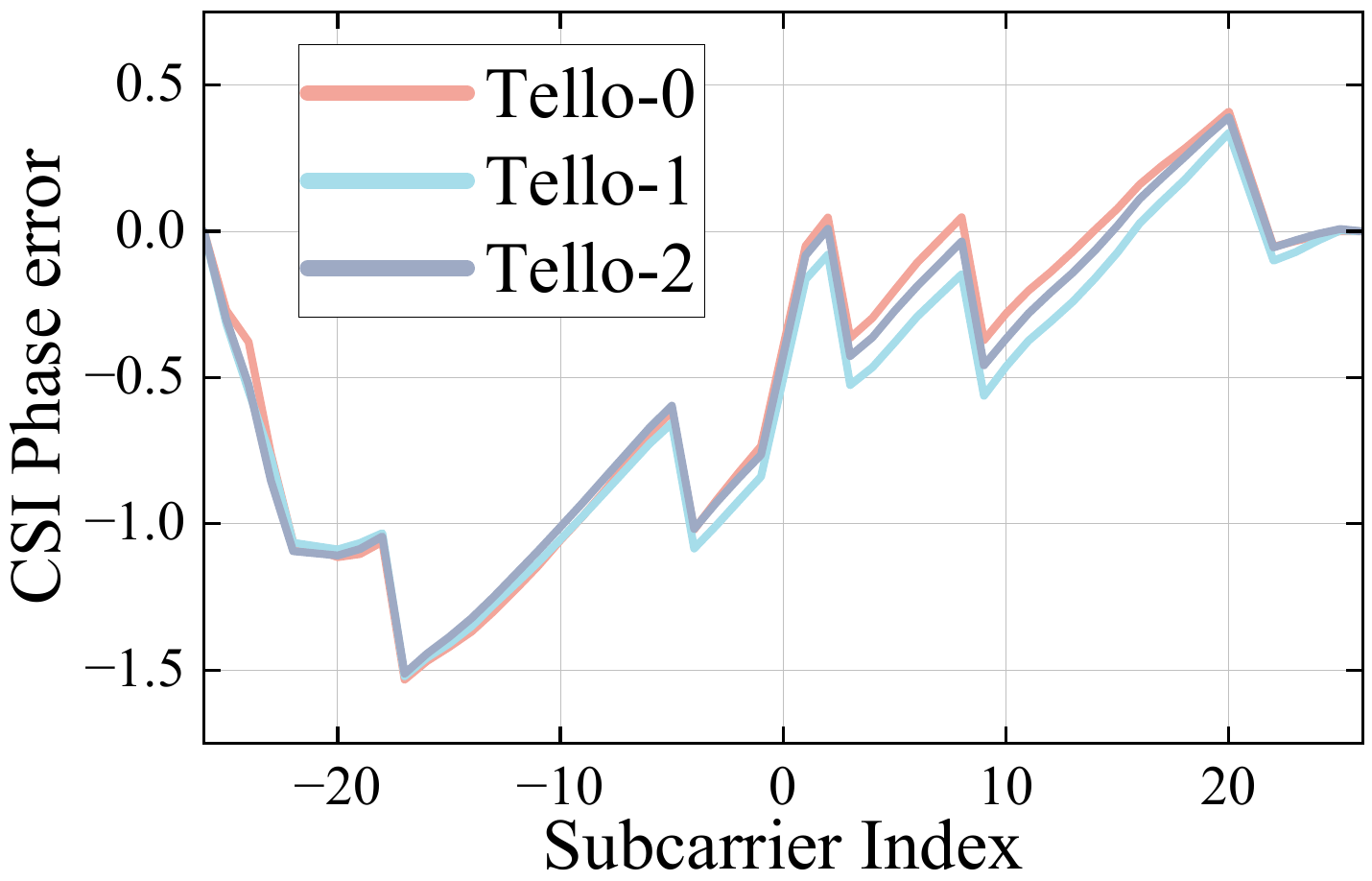}
     \label{fig:Tello_CSI_phase_errors_dynamic}
     }
    \subfigure[Accelerometer X-axis outputs.]{
    	\includegraphics[width=0.46\columnwidth]{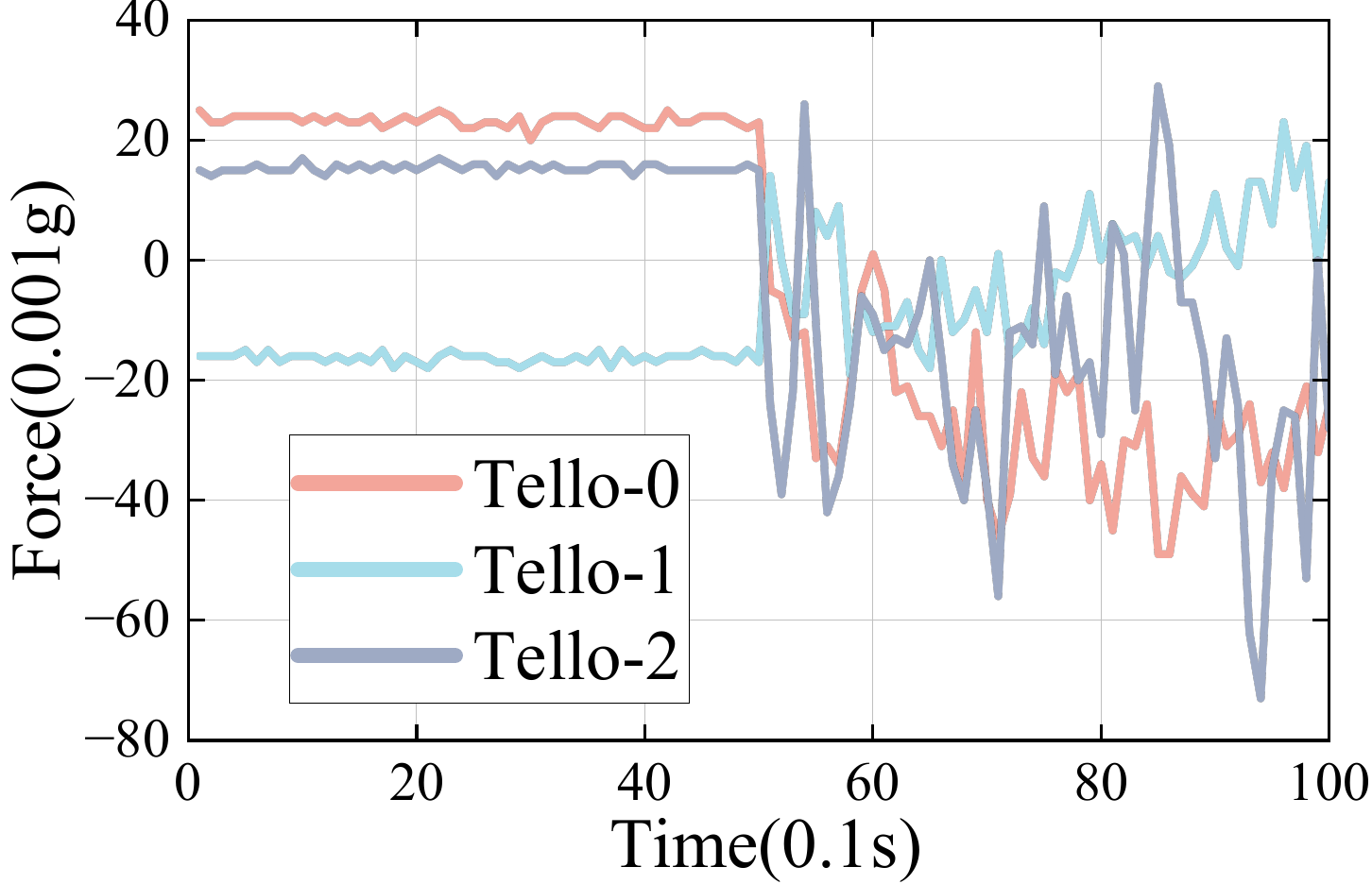}
    	\label{fig:Tello_acc_x}
    }
    \subfigure[Accelerometer Y-axis outputs.]{
    	\includegraphics[width=0.46\columnwidth]{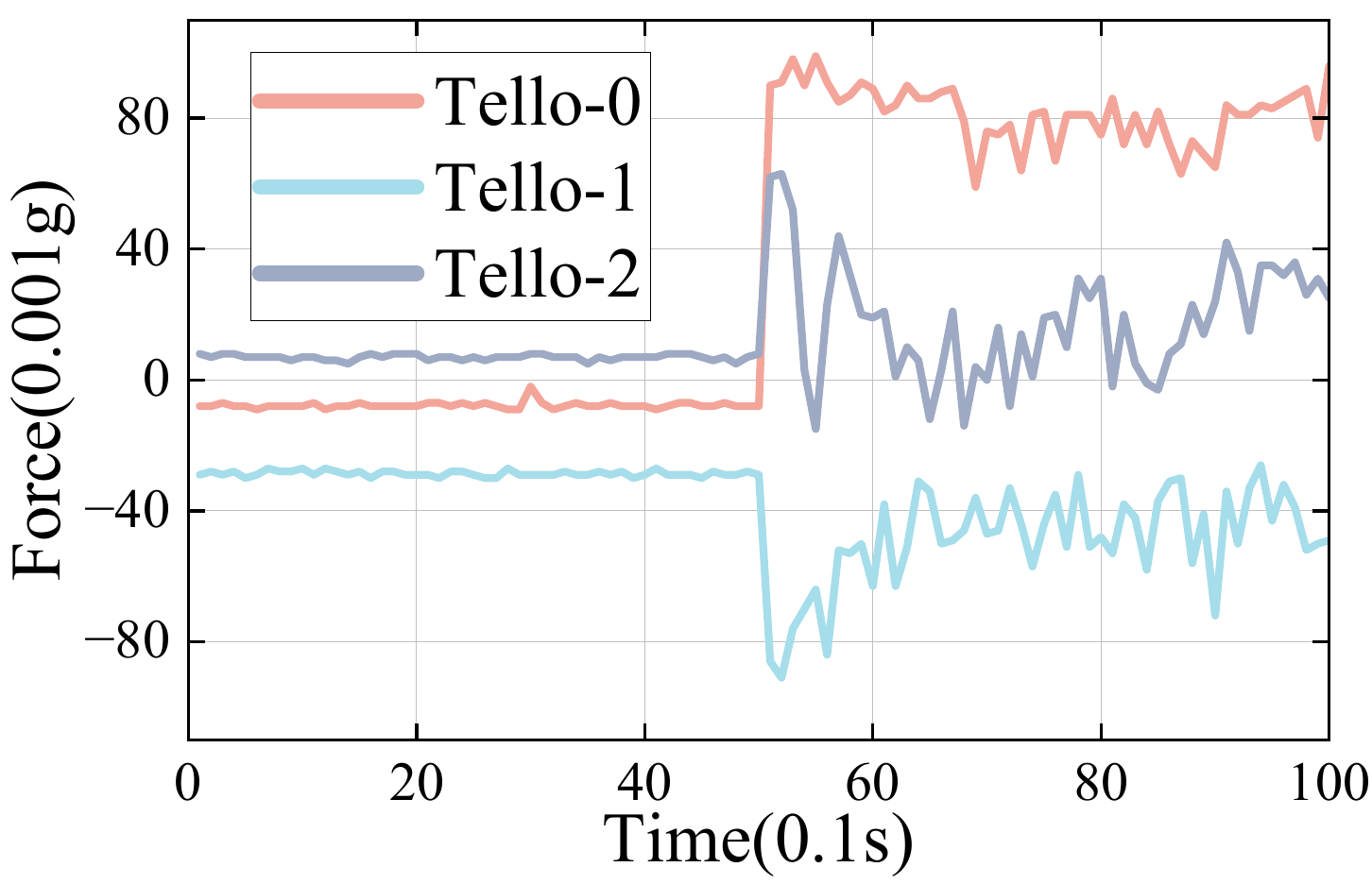}
    	\label{fig:Tello_acc_y}
    }
    \caption{RF and MEMS fingerprints in different motion states.} 
    \label{fig:Tello_fingerprints}
\end{figure}

\subsection{Feasibility Study}\label{subsec:empirical_experiment}

We conduct preliminary experiments to verify the feasibility of discriminating commercial-off-the-shelf UAVs using CSI and telemetry measurements.
For this purpose, we take a Legion Y9000X laptop as a GCS to control three DJI Tello drones.
The laptop runs PicoScenes~\cite{picoscenes} and the Tello software development kit (SDK) to simultaneously collect CSI and telemetry readings from the three UAVs.
During data collection, we control each UAV to stay on the ground in the first five seconds and then keep it in the flying state in the next five seconds.
The CSI phase errors and accelerometer outputs on the X and Y axes are present in Fig.~\ref{fig:Tello_fingerprints}.

As Fig.~\ref{fig:Tello_CSI_phase_errors_static} and~\ref{fig:Tello_CSI_phase_errors_dynamic} show, the phase errors of the three UAVs are very close, and some error features even overlap in each state. 
This is because all UAVs are the same make and their RF transceivers share similar hardware imperfections.
Despite that, we can still observe some minor differences in some subcarriers, showing the potential of CSI-based UAV fingerprinting.
In addition, the phase errors of one UAV are slightly different in the stationary and flying states.
Such differences are caused by the phase drift~\cite{PEDR}, which is incurred by environmental dynamics.
As depicted in Fig.~\ref{fig:Tello_acc_x} and Fig.~\ref{fig:Tello_acc_y}, the accelerometer outputs of three UAVs are stable in the first five seconds. 
This is because there is only the bias error in the stationary state according to Eq.~\eqref{eq:sensors error}.
Although the accelerometer outputs fluctuate dramatically in the flying state, different changing patterns can be observed among the three UAVs. 
In conclusion, though airborne RF and MEMS imperfections are sensitive to environmental dynamics and UAV mobility, they are still distinguishable between different commercial-off-the-shelf UAVs.

\section{System Design}\label{sec:design}

\subsection{System Overview}

We propose SecureLink, a novel cross-layer UAV authentication system that can accurately authenticate commercial-off-the-shelf UAVs using onboard RF and MEMS fingerprints. 
SecureLink runs on the GCS, a ground-based computer processing unit with relatively abundant computational resources. 
SecureLink does not have computational requirements for UAVs.
In practice, SecureLink works as follows.
Initially, a UAV sends an authentication request with a claimed ID and a certificate to the GCS.
Once the authentication passes the cryptography-based check of the GCS, the telemetry data transmission between the UAV and the GCS starts.
Then, the GCS triggers SecureLink to employ the CSI and sensory readings from telemetry data to conduct a secondary UAV verification. 
If the extracted features match the fingerprints of the claimed UAV, the authentication request is authorized.
Otherwise, the request is denied.
In this way, SecureLink follows the cryptography-based authentication process and relies on downlink telemetry data packets for UAV verification, and has no impact on the command and payload traffic.
Thus, SecureLink introduces limited communication overheads.

\begin{figure}
	\centering
	\includegraphics[width=1.0\linewidth]{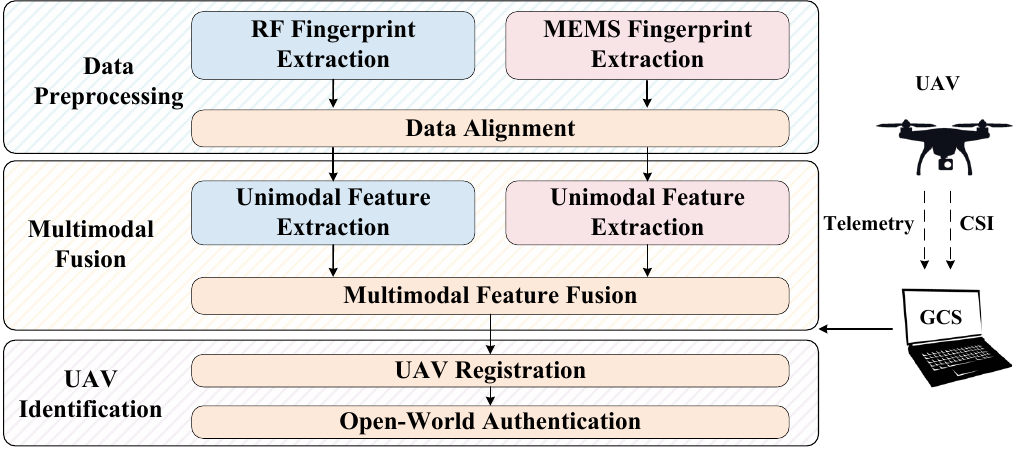}
	\caption{System architecture of SecureLink. It mainly consists of data preprocessing, multimodal fusion, and UAV identification.}
	\label{fig:systemflow}
\end{figure}

As depicted in Fig.~\ref{fig:systemflow}, the core of SecureLink includes the following three components.
\begin{itemize}
    \item
    \textbf{Data Preprocessing.}
    This component first extracts phase errors from raw CSI measurements and selects useful fields from telemetry data.
    Then, it filters out outliers from the two types of fingerprints.
    Next, this component aligns CSI fingerprints to telemetry fingerprints using low-complexity linear interpolation.
    \item
    \textbf{Multimodal Fusion.} 
    This component first utilizes two neural network branches, with a 1D CNN and a BiLSTM each, to extract unimodal features from two modalities, respectively.
    Then, the unimodal features are concatenated and fed into two multi-head attention layers to generate multimodal vectors. 
        \item
    \textbf{UAV Identification.}
    This component first trains a classification model using a multi-similarity loss for UAV registration.
    Then, it utilizes OC-SVM to estimate the likelihood of unknown UAVs and thus empowers open-world UAV authentication.

\end{itemize}

\subsection{Data Preprocessing}

\textbf{RF Fingerprint Extraction.} 
CSI describes the propagation characteristics of wireless signals and is susceptible to environmental dynamics.
Hence, it is necessary to extract robust fingerprinting features from raw CSI measurements.
For this reason, we exploit the CSI phase errors as our RF fingerprints due to their relative resilience to environment changes compared to CSI amplitudes~\cite{CSI-Phase-error}.
The phase errors can be extracted as follows. 
First, we calculate the phases of all subcarriers and unwrap each phase.
Then, the zero subcarrier, also called the direct component, is removed.
Next, to remove outliers caused by environmental interference, the phase frames with steep variance are filtered out.
Let us denote $\mathbf{\phi} = \left( \phi_1, \cdots, \phi_k, \cdots, \phi_K \right)$ as the phases of one CSI measurement, where $K$ is the number of CSI subcarriers.
The phase gradient $\nabla{\Phi}$ can be calculated as $\nabla{\Phi} = \left( \phi_2 - \phi_1, \cdots, \phi_k -\phi_{k-1}, \cdots, \phi_K - \phi_{K-1} \right)$.
If the variance of $\nabla{\Phi}$ is smaller than the threshold $\eta$, the phase vector $\mathbf{\phi}$ is reserved.
After that, we estimate the values of $\lambda$ and $\mathbf{z}$ using the mirror subcarriers~\cite{CSI-Phase-error}.
Consequently, the phase error $\mathbf{e}$ of $\mathbf{\phi}$ can be calculated and denoted as $\mathbf{e} = \left( e_1, \cdots, e_k,\cdots, e_K \right)$.
During data collection, we obtain a CSI fingerprint sequence as $\mathbf{E} = \left\lbrace \mathbf{e}^1, \cdots, \mathbf{e}^l,\cdots, \mathbf{e}^L \right\rbrace$, where $ \mathbf{e}^{l} $ is the $l$-th CSI fingerprint frame and $L$ indicates the total number of CSI measurements.

\begin{figure}
    \centering
    \subfigure[ToF measurements and corresponding outliers.]{
    	\includegraphics[width=0.46\columnwidth]{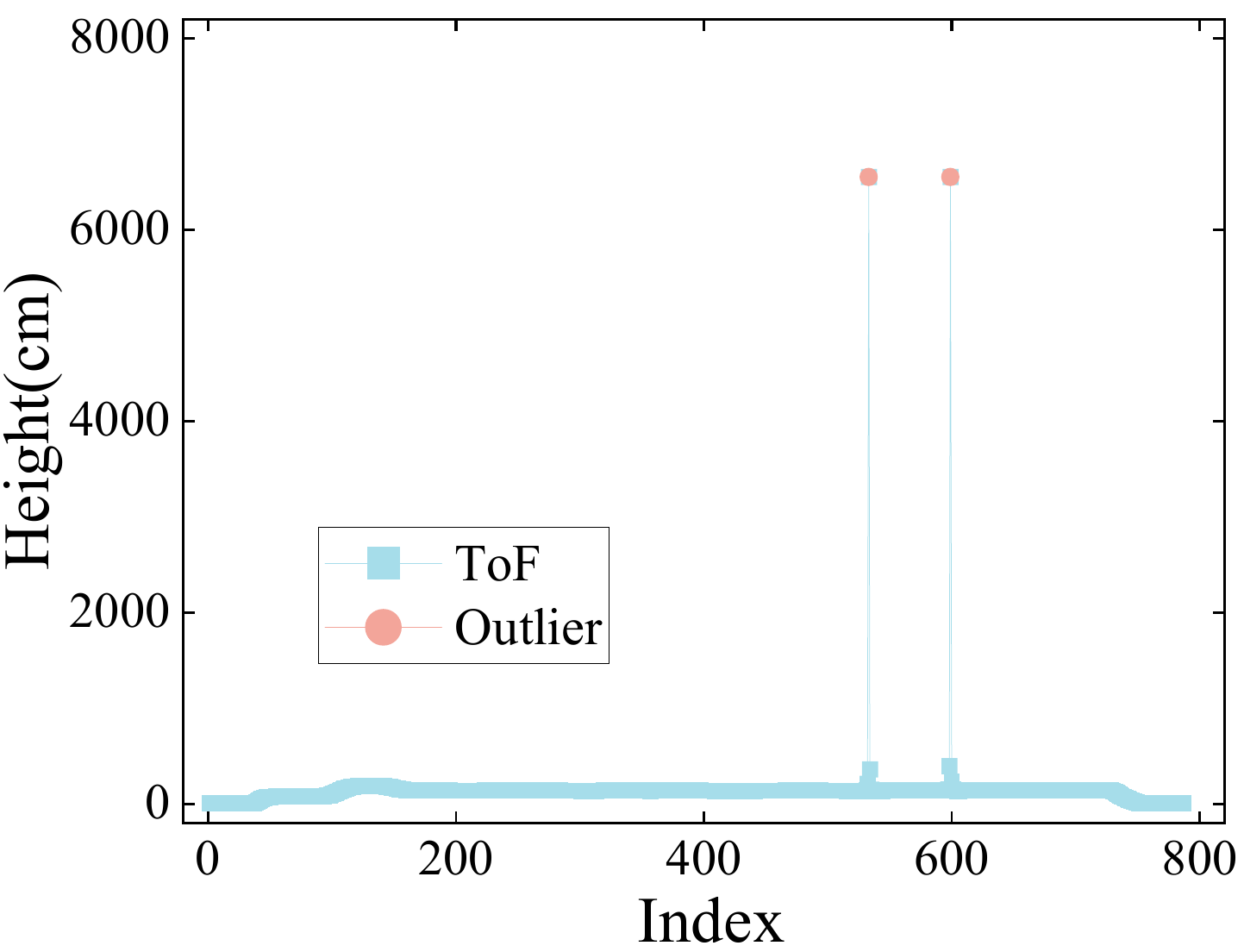}
    	\label{fig:tof_outlier}
     }
     \subfigure[Time intervals between successive CSI measurements.]{
     \includegraphics[width=0.46\columnwidth]{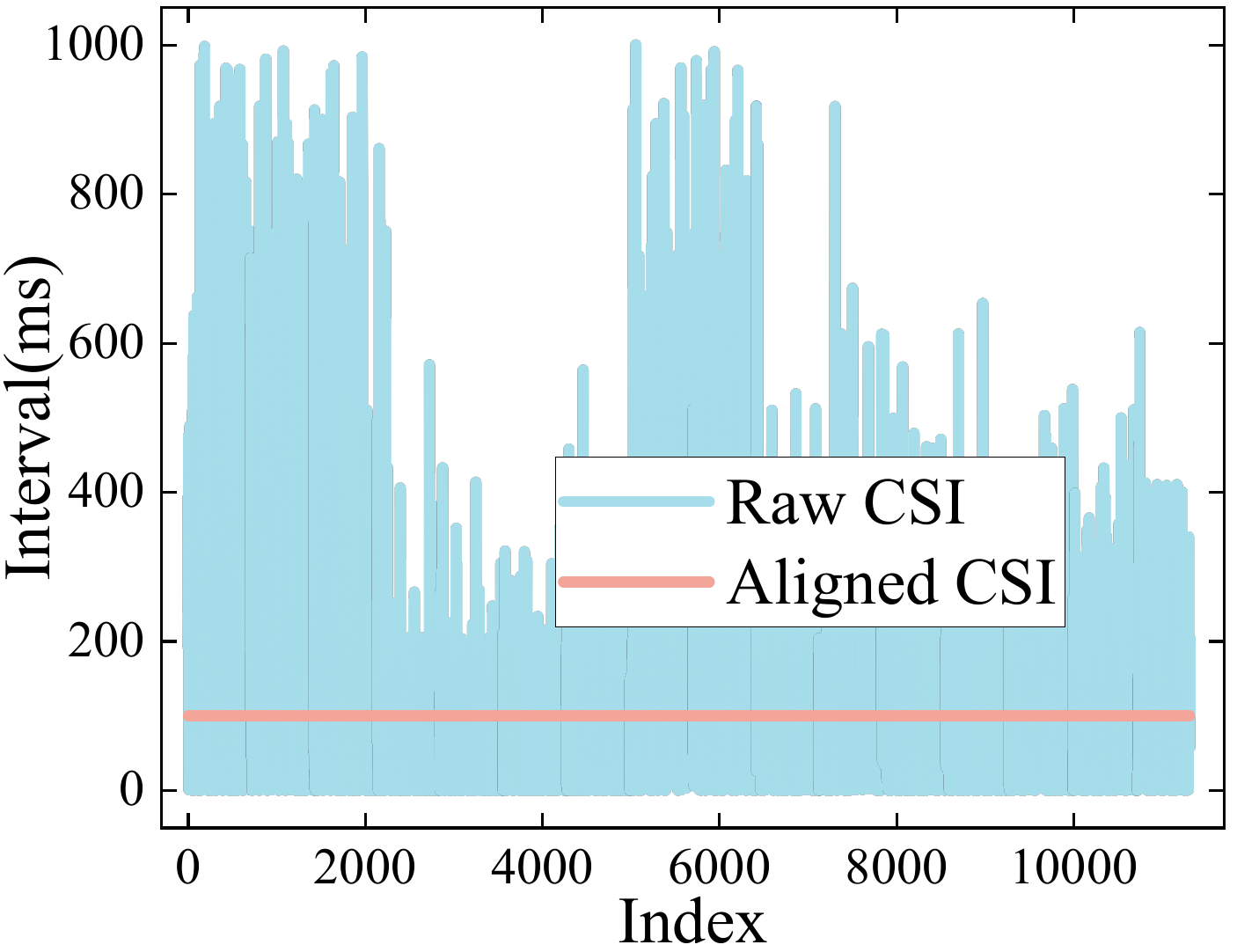}
     \label{fig:aligned_csi}
     }
    \caption{ToF and CSI measurements.} 
    \label{fig:data_preprocessing}
\end{figure}

\textbf{MEMS Fingerprint Extraction.}
We select useful sensory readings as MEMS fingerprints.
Specifically, the telemetry data contains hundreds of fields about UAV status, such as heartbeat, system parameters, rotor status, memory status, global positioning system (GPS) data, sensor outputs, and so on.
Because we are only interested in a few MEMS-sensor-related information, we select a total of eight fields, consisting of pitch, roll, yaw, and outputs of the three-axis accelerometer, barometer, and ToF sensor.
In this way, one telemetry fingerprint frame can be denoted as $ \mathbf{s} = \left(s_1, s_2, \cdots, s_8 \right) $.
Moreover, during UAV flight, there could be outliers in some fields due to environmental noise or random failure, leading to field values out of the measurement range.
For example, the ToF outputs are set to be the maximum value of the register when the UAV flips in the air, as depicted in Fig.~\ref{fig:data_preprocessing}~(a).
To avoid the detrimental effects of such abnormal values, we clean up all telemetry frames containing outliers.
Fortunately, such outliers are quite sparse in all sensor measurements.
During data collection, we can obtain a telemetry fingerprint sequence as $\mathbf{S} = \left\lbrace \mathbf{s}^1,\cdots, \mathbf{s}^m,\cdots,  \mathbf{s}^M \right\rbrace$, where $\mathbf{s}^m$ is the $m$-th telemetry frame and $M$ indicates the total number of frames. 

\textbf{Data Alignment.}
Multimodal fusion requires the alignment of different modalities for discovering underlying correlations.
Generally, CSI is sensitive to environmental dynamics, resulting in many filtered CSI measurements.
In contrast, the occurrence probability of outliers in MEMS sensor measurements is much lower.
Under these conditions, the number of CSI fingerprint frames is much smaller than that of telemetry frames during data collection, i.e., $L \ll M$.
To deal with this issue, we align the two modalities using a linear interpolation method.
Since $L < M$, we align the two types of frames based on their timestamps and resample $M-L$ new CSI frames from raw frames using low-complexity linear interpolation.
The above process can be denoted as
\begin{align}
    \mathbf{E}' = \text{Interpolation} (\mathbf{E}, M-L) = \left\lbrace \mathbf{e}^1,\cdots, \mathbf{e}^m,\cdots,  \mathbf{e}^M \right\rbrace.
\end{align}
 After the interpolation, the frame number of $\mathbf{E}$ extends to $M$, and $\mathbf{E}$ is synchronized with $\mathbf{S}$ as shown in Fig.~\ref{fig:data_preprocessing}~(b).

\subsection{Multimodal Fusion}
\begin{figure}
    \centering
	\includegraphics[width=1.0\linewidth]{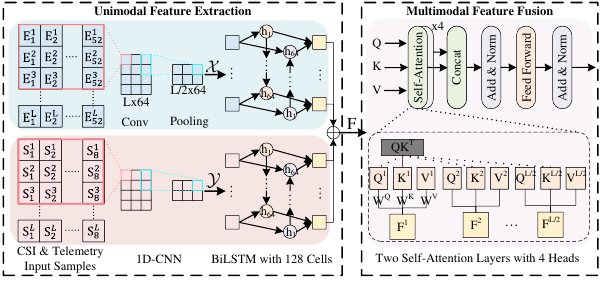}
	\caption{The workflow of multimodal fusion. It has unimodal feature extraction and multimodal feature fusion.}
	\label{fig:multimodal_fusion}
\end{figure}

Traditional UAV authentication approaches rely on a single modality of sensory data, falling short in robustness and efficiency.
Hence, we choose to fuse the RF and MEMS fingerprints into high-dimensional features.
As shown in Fig.~\ref{fig:multimodal_fusion}, we first devise two neural network branches to extract unimodal features from each modality.
Next, the extracted unimodal features are concatenated and passed to two multi-head attention layers for adaptive feature fusion.

\textbf{Unimodal Feature Extraction.} 
As indicated in the feasibility study, CSI phase errors are impacted by environmental dynamics, and telemetry features are sensitive to UAV motion changes, rendering hand-crafted features ill-suited to this task.
Inspired by the recent achievements in deep learning, we devise two neural network branches with the same structure to extract unimodal features from raw CSI and telemetry fingerprints, respectively.

Specifically, each branch has a convolutional neural network followed by a bidirectional long short-term memory layer.
Therein, one CNN consists of two consecutive 1D convolutional layers with a pooling layer.
In its first convolutional layers, 64 1D convolutional kernels with a size of 3 extract high-level features from one modality.
After that, the extracted features are input to a pooling layer, followed by the second convolutional layer with 64 kernels in a size of 2.
In this CNN, rectified linear units (ReLUs) are used as activation functions to introduce non-linearity, and batch normalization layers are configured after the input and the convolutional layers to reduce the risk of overfitting.
Then, the CNN features are fed into a BiLSTM for feature enhancement.
Because CSI and telemetry fingerprints would vary in different environments and UAV motions, it is critical to extract persistent fingerprinting information across different data frames. 
BiLSTM can analyze the temporal evolution of these frames to capture more potentially useful information, thus reducing the impact of environmental and motion changes on feature extraction.
For this reason, we adopt a BiLSTM layer with 128 cells to aggregate useful information in each fingerprint sample from its forward and backward directions.

Taking a pair of aligned fingerprint sequences $\mathbf{E}'$ and $\mathbf{S}$ as input, the feature extraction component outputs the unimodal feature vectors of the two modalities as
\begin{align}
\mathbf{X} &= \mathcal{F}_{UF} (\mathbf{E}') = \left\lbrace \mathbf{x}^1, \cdots, \mathbf{x}^m, \cdots, \mathbf{x}^{M/2} \right\rbrace, \\
\mathbf{Y} &= \mathcal{F}_{UF} (\mathbf{S}) = \left\lbrace \mathbf{y}^1, \cdots, \mathbf{y}^m, \cdots, \mathbf{y}^{M/2}\right\rbrace,
\end{align}
where $\mathbf{x}^m \in \mathbb{R}^{128}$ and $\mathbf{y}^m \in \mathbb{R}^{128}$ are the unimodal features of CSI and telemetry fingerprint samples, respectively.
$\mathcal{F}_{UF}(\cdot)$ denotes the designed network for unimodal feature extraction.
Finally, we concatenate the two unimodal features after the BiLSTM layer and obtain a feature map as $\mathbf{U} = \mathbf{X} \oplus \mathbf{Y}$, where $\oplus$ denotes the concatenation operator.

\textbf{Multimodal Feature Fusion.}
UAVs' RF and MEMS fingerprints would typically vary in different working environments and UAV motions.
Hence, it is unwise to fuse them by simply concatenating two unimodal features and assigning each feature value the same weight in a feature map $\mathbf{U}$. 
Thus, an adaptive feature fusion scheme is desirable.

To achieve this, we perform feature fusion based on the multi-head attention mechanism.
The multi-head attention scheme was proposed by Vaswani et al.~\cite{transformer} and subsequently widely applied to many deep learning tasks.
This mechanism captures features in different subspaces through multiple independent attention heads.
Each head can learn distinct feature representations, allowing our model to understand and represent the RF and MEMS fingerprints from various perspectives.
Specifically, we implement our model with four attention heads. 
The calculation process can be written by
\begin{align}
    \text{MultiHead}(\mathbf{Q}, \mathbf{K}, \mathbf{V} ) &= (\text{head}_1 \oplus \cdots \oplus \text{head}_4)\mathbf{w}^o, \\
    \text{head}_i &= \text{Softmax}\left(\frac{\mathbf{q}_{i}\mathbf{k}^T_{i}}{\sqrt{d_i}}\right)\mathbf{v}_{i},
\end{align}
where $i=1,\cdots, 4$ and $\mathbf{w}^o$ is the weight matrix of the linear transformation.
$\mathbf{Q},\mathbf{K},\mathbf{V}$ denote attentional weights of different subspaces: Query, Key, and Value, respectively.
The query $\mathbf{Q}$ represents the information of interest in input feature vectors.
The key $\mathbf{K}$ indicates the position of each piece of such information.
The value $\mathbf{V}$ represents various information contained in identity-related features.
Moreover, $\mathbf{q}_i$, $\mathbf{k}_i$, and $\mathbf{v}_i$ are the results of the dot product of input vectors and weight matrices of three subspaces in the $i$-th head.
$d_i$ is the dimension of the input data.
Based on the above equations, multi-head attention can not only capture the relevance of different parts of unimodal features but also highlight the parts most relevant to UAV identities by learning attentional weights.
With the multi-head attention layer, higher weights can be adaptively assigned to the important parts of the concatenated feature maps, thus improving the ability of feature fusion for UAV fingerprints.

After two multi-head attention layers, we proceed to flatten its output into a one-dimensional multimodal feature vector.
Finally, given a feature map $\mathbf{U}$, we obtain a multimodal feature vector $\mathbf{D}$ as $\mathbf{D} = \mathcal{F}_{MF} \left(\mathbf{U} \right)$, where $\mathcal{F}_{MF} (\cdot) $ represents the neural network parameters of multimodal fusion.

\subsection{UAV Identification}

\textbf{UAV Registration.}
After the multimodal fusion, we need to register the multimodal fingerprints of $H$ authorized UAVs before authentication.
We aim to ensure that the fingerprint similarity between different UAVs is minimized, while the similarity among fingerprints of the same UAV is maximized.

To achieve this, a metric embedding layer is exploited to convert each multimodal vector into a lower-dimensional embedding space where distances between input data points reflect identity similarity.
Specifically, we build this using a fully connected layer with a size of 256, which compresses the high-dimensional multimodal vector into a compact vector and learns meaningful features that capture the similarity or dissimilarity between multimodal vectors.
The output embeddings can be expressed as $\mathbf{g} = \mathcal{F}_{EB} \left(\mathbf{D}\right)$, where $\mathcal{F}_{EB} (\cdot) $ represents the parameters of the embedding layer.

In the training phase, we adopt a multi-similarity loss~\cite{wang2019multi} to train our neural networks $\mathcal{F}_{UF} (\cdot)$, $\mathcal{F}_{MF}(\cdot)$ and $\mathcal{F}_{EB} (\cdot) $.
Its primary goal is to optimize these networks to learn an effective embedding space where similar samples are closer together, and dissimilar samples are farther apart. 
Concretely, a sample is randomly selected as the anchor, paired with other samples from the same UAV to form positive pairs.
Correspondingly, the anchor is paired with the samples from different UAVs to construct negative pairs.
The multi-similarity loss captures the self-similarity, positive similarity, and negative similarity of the sample pairs within the training dataset, assigning different weights to each.
Thus, we can train our model by minimizing the multi-similarity loss $\mathcal{L}_{MS}$, which is computed as
\begin{align}\label{eq: loss}
    \mathcal{L}_{MS}  =  \frac{1}{A} \sum_{i=1}^A \Bigg[ \frac{1}{\alpha} \log \left( 1 + \sum_{j \in \mathcal{P}_i} \exp{(-\alpha (C_{ij} - \mu))} \right) \nonumber \\
    + \frac{1}{\beta} \log \left( 1 + \sum_{r \in \mathcal{N}_i} \exp{(\beta (C_{ir} - \mu))} \right) \Bigg],
\end{align}
where $ i $ represents the $i$-th anchor sample, and $A$ is the number of all anchor samples in the batch.
$ \mathcal{P}_i $ and $ \mathcal{N}_i $ are the sets of positive and negative samples for the $i$-th anchor,
$ C_{ij} $ and $ C_{ir} $ represent the anchor's similarity scores relative to the $j$-th positive sample and the $r$-th negative sample, respectively. 
Therein, the similarity between sample pairs is calculated using the cosine similarity of their embedding vectors.
Moreover, $ \mu $ is the similarity margin, which helps control the influence of samples near the decision boundary, 
and $ \alpha $ and $ \beta $ are scaling parameters.
In Eq.~\eqref{eq: loss}, the first part makes the fingerprints of the same UAV closer, and the second part forces those of different UAVs farther.
After training, the multimodal fingerprints of the $H$ registered UAVs are obtained.
In the testing phase, the trained neural networks can be leveraged to generate new UAV fingerprints. 

\textbf{UAV Authentication.}
After UAV registration, our model can identify which registered UAV the test fingerprint corresponds to by selecting the UAV ID with the highest similarity score.
However, this may mistakenly classify an unseen impersonating UAV as one of the registered ones.
To achieve open-world authentication, we use a OC-SVM~\cite{ocsvm} to check whether the connecting UAV is legitimate.

Generally, the standard SVM is a binary classifier that discriminates between two distinct classes by building a decision hyperplane. 
However, this binary approach is unsuitable for one-class classification problems, where only positive examples are available. 
In contrast, the OC-SVM serves as a unary classifier, capable of constructing a support vector representation solely based on positive samples. 
Formally, let $ \left \lbrace \mathbf{g}_h^1 , \cdots, \mathbf{g}_h^b ,\cdots, \mathbf{g}_h^B \right \rbrace $ be the embeddings belonging to the $h$-th registered UAV.
In the OC-SVM, these embeddings are mapped to a higher-dimensional feature space using a Gaussian kernel function $\zeta(\cdot)$.
In this space, OC-SVM seeks a hyperplane that maximizes the margin from the origin while encircling most of the training samples.
The training process can be formulated as a quadratic programming problem as 
\begin{align}
\min_{\mathbf{n}, \rho, \xi} \frac{1}{2} \|\mathbf{n}\|^2 + \frac{1}{\tau B} \sum_{b=1}^B \xi_b - \rho,
\end{align}
\begin{align}
\text{s.t.} \;\; \mathbf{n} \cdot \zeta (\mathbf{g}_h^b) \geq \rho - \xi_b, \quad \xi_b \geq 0, \quad b = 1, \cdots, B,
\end{align}
where $\mathbf{n}$ and $\rho$ are a weight vector and an offset parameterizing a hyperplane in the feature space associated with the Gaussian kernel.
$\xi_b$ is the slack variable to allow for soft margins, \( \tau \in [0, 1] \) is a parameter that controls the trade-off between the regularization term and the margin violations.
If $\mathbf{n}$ and $\rho$ solve the optimization problem, the decision hyperplane can be represented by $\mathcal{F}_{OCSVM}(\mathbf{g}_h) = \text{Sign}(\mathbf{n} \cdot \zeta(\mathbf{g}_h) - \rho)$, where $\text{Sign} (\cdot)$ denotes the sign function.
$\mathcal{F}_{OCSVM} (\cdot)$ will be positive for most true fingerprints while allowing some deviations controlled by $ \tau $ and $ \xi $.
Though a OC-SVM is trained for each UAV in the training phase, only the OC-SVM belonging to the claimed ID is used in the inference phase.
Thus, no excessive computation overheads are generated.

Given a pair of new CSI and MEMS features, our proposed neural networks $\mathcal{F}_{UF} (\cdot)$, $\mathcal{F}_{MF}(\cdot)$, and $\mathcal{F}_{EB} (\cdot) $ can be leveraged to generate a multimodal fingerprint, which is subsequently fed into the decision function $\mathcal{F}_{OCSVM} (\cdot)$ corresponding to the claimed UAV ID.
If the fingerprint falls in the decision hyperplane, the connecting UAV is considered to be legitimate.
Otherwise, an impersonating UAV is detected, and the association is denied by the GCS.
Thus, our system can authenticate legitimate UAVs and detect illegal ones simultaneously.

\begin{figure}[t]
	\centering
	\includegraphics[width=1.0\linewidth]{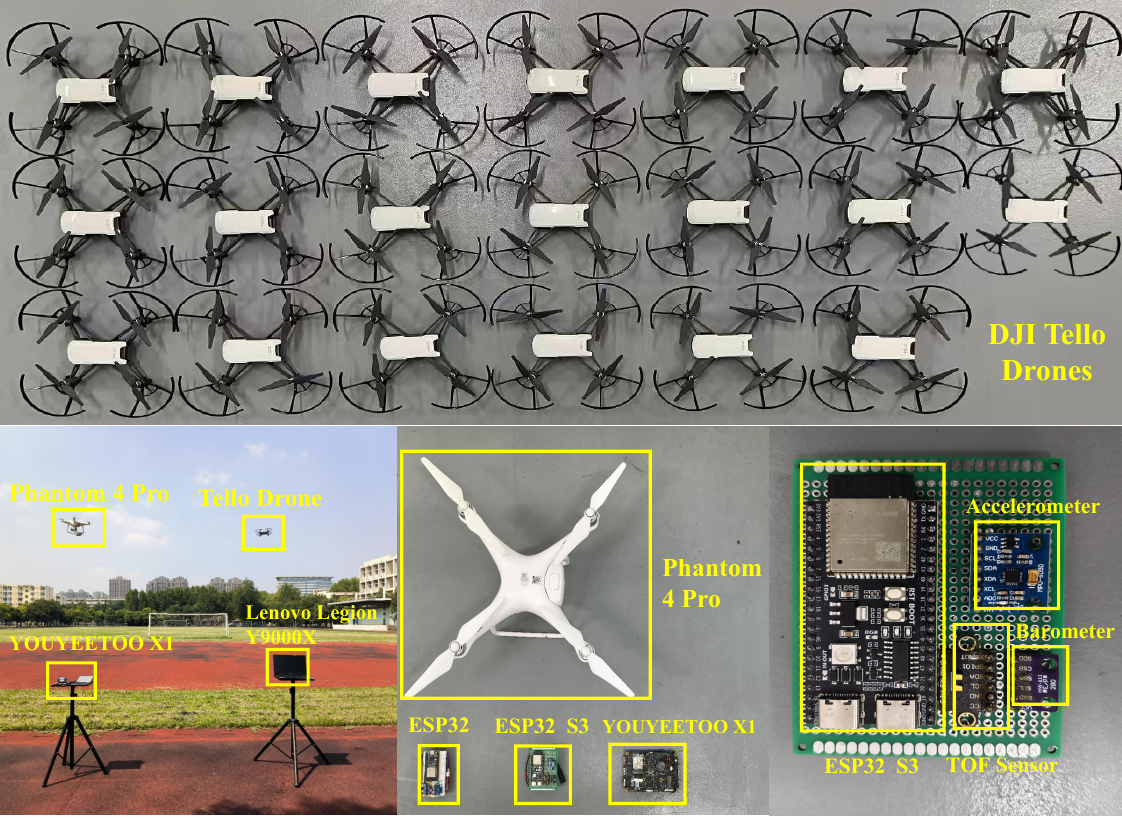}
	\caption{The experimental platform with three types of drones.}
	\label{fig:experimental_setting}
\end{figure}

\section{Implementation and Evaluation}\label{sec:implementation}

\subsection{Implementation}
As shown in Fig.~\ref{fig:experimental_setting}, we develop the following three types of UAVs in our testbed to simulate data collection and transmission conducted by heterogeneous UAV models.
\begin{itemize}
    \item \textbf{DJI Tello.} There are twenty DJI Tello drones in our testbed. Each Tello drone is a quadrotor and has a built-in 802.11n 2.4GHz Wi-Fi module with a single antenna. 
    We use a Lenovo Legion Y9000X laptop as the GCS, which has an Intel Wi-Fi 6E AX211 NIC and an Intel Core i7-12700H processor with a 16GB-RAM and runs on Ubuntu 22.04 LTS.
    The laptop uses the \verb|"tello_test.py"| in the DJI Tello SDK to control each UAV and the \verb|"tello_state.py"| script to receive and store Tello telemetry readings.
    In addition, PicoScenes~\cite{picoscenes} is leveraged to collect Wi-Fi CSI measurements of 52 subcarriers from Tello drones.
    \item \textbf{DJI Phantom 4 Pro with ESP32.} We build a UAV prototype using a DJI Phantom 4 Pro drone with an ESP32 SoC, which supports IEEE 802.11b/g/n WiFi communications and is powered by a 5V DC battery. 
    The SoC is equipped with an MPU6050 accelerometer, a BMP280 barometer, and a VL53L4CD ToF sensor.
    As for the GCS, it utilises an 8GB-RAM YOUYEETOO X1 development board and an Intel WiFi 6E AX201 NIC, and runs Ubuntu 22.04 LTS with the DJI Tello SDK.
    We use Arduino IDE to write firmware into the SoC and facilitate data transmissions with the YOUYEETOO X1.
    In addition, the SoC is carried by the DJI Phantom 4 Pro to collect and transmit data during flight.
    \item \textbf{DJI Phantom 4 Pro with ESP32-S3.} We also build another UAV prototype using a DJI Phantom 4 Pro drone with an ESP32-S3 SoC is developed.
    The SoC has an MPU6050 accelerometer, a BMP280 barometer, and a VL53L1X ToF sensor.
    The other specifications and configurations are the same as those of the above prototype.
\end{itemize}
The hardware specifications of different types of UAVs are summarized in Table~\ref{tab: drone hardware}.
In this way, we have 22 UAVs of different models and assign each UAV a unique number, i.e., an ID. 
The Tello drones are marked with numbers, ranging from 1 to 20, and the DJI Phantom 4 Pro with the ESP32 SoC and the ESP32-S3 SoC are designated to 21 and 22, respectively.

\begin{table}[t]
\centering
\caption{Hardware Specifications of Different Types of UAVs}
\label{tab: drone hardware}
\begin{tabular}{cccc}
\hline
      & DJI Tello                                                  & \begin{tabular}[c]{@{}c@{}}DJI Phantom 4 Pro \\ with ESP32\end{tabular} & \begin{tabular}[c]{@{}c@{}}DJI Phantom 4 Pro \\ with ESP32-S3\end{tabular} \\ \hline
Acc.  & LSM6DS3TR-C                                                & MPU6050                                                                     & MPU6050                                                                        \\
Baro. & MS5611-01BA03                                              & BMP280                                                                      & BMP280                                                                         \\
ToF   & VL53L0X                                                    & VL53L4CD                                                                    & VL53L1X                                                                        \\
NIC   & Marvell 88W8801  & D0WDQ6                            & D0WDQ6                                                                         \\ \hline
\end{tabular}
\end{table}

\subsection{Evaluation Methodology}

\textbf{Data Collection.} 
First, we collect multimodal data in the dynamic state in four environments, including a rooftop, a playground, a corridor, and an office.
The data collection procedure is as follows.
At first, we set each UAV into the SDK mode by sending corresponding commands from the laptop.
In this mode, the UAV sends telemetry readings to the GCS at a frequency of 10Hz.
Next, we run the two scripts to collect telemetry data and the PicoScenes tool to record Wi-Fi CSI measurements at the same time.
Then, we send a sequence of flight commands to the UAV to perform scheduled flight actions, such as up, down, left, right, rotate, and so on.
This command sequence puts each UAV in the flying state for about 100 seconds.
Finally, 22 UAVs take turns to complete the above procedure and repeat it five or six times in each environment.
Second, we collect data in the stationary state in the office.
In this scenario, each UAV is turned on, but its motor is not activated. 
Likewise, all UAVs repeat the above procedure five times each.
After all, we collect multimodal data for more than twelve hours on different days.
The collected data are available on GitHub~\cite{dataset}.

\textbf{Datasets.}
After data collection, we perform data preprocessing on the collected telemetry and CSI data, and obtain more than 400K aligned data frames, with approximately 20K samples for each UAV.
For effective training and testing, we split all samples into two datasets.
\begin{itemize}
    \item \textbf{Dataset A.}
    It is for closed-world evaluation.
    Each sample is labeled with its true UAV ID.
    All samples are used in the training and testing phases.
    For each UAV, we shuffle all samples and split them into 60\%, 20\%, and 20\% for training, validation, and testing.
    \item \textbf{Dataset B.} 
    It is for open-world evaluation.
    Twenty UAVs are selected as registered ones.
    Their samples are labeled with their true UAV IDs.
    The other two are impersonating others, and their samples are assigned a registered ID randomly.
    80\% of samples from the registered UAVs are used for training and validation.
    The remaining 20\% of samples from the registered UAVs and all samples from the illegitimate ones are used for testing.
\end{itemize}

\textbf{Model Training.}
We build our system using Python 3.9.19 and TensorFlow 2.13.1 and train it on an Omnisky AIX7550-G3 workstation with an NVIDIA RTX A5000 GPU chip.
During training, the learning rate is set to 0.001, the batch size to 256, and the training epochs to 30.
Moreover, we adopt the Adam optimizer for parameter learning.
In addition, the multimodal fingerprints of each UAV produced by the trained neural networks are used to train OC-SVMs.

\textbf{Evaluation Metrics.}
We use the accuracy, the true negative rate, the recall, and the precision as our evaluation metrics.
\begin{itemize}
    \item \textbf{Accuracy.} It is defined as the ratio of all correctly classified feature samples to all samples.
    \item \textbf{TNR.} It is the proportion of unauthorized samples that are correctly rejected by the system. 
    \item \textbf{Recall.} It is defined as the ratio of correctly authorized samples to the total number of legitimate samples.
    \item \textbf{Precision.} It is the ratio of correctly authorized samples to the total number of samples classified as legitimate.    
\end{itemize}

\begin{figure}
    \centering
	\includegraphics[width=1.0\linewidth]{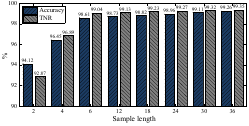}
	\caption{Performance under different sample lengths on dataset A.}
	\label{fig:parameter_determination}
\end{figure}

\begin{figure}[t]
	\centering
	\includegraphics[width=1.0\linewidth]{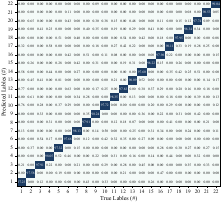}
        \caption{Confusion matrix in the closed world. \#1 to \#20 stand for the Tello drones, and \#21 and \#22 are the DJI Phantom 4 Pro with the ESP32 SoC and the ESP32-S3 SoC, respectively.}
	\label{fig:closed-world_performance}
\end{figure}

\subsection{Experimental Results.}

\textbf{Closed-World Performance.}
The first step of our experiment is to decide the sample length $M$, i.e., the number of multimodal data frames contained in each fingerprint sample.
Generally, the larger the sample length is, the more information will be provided and thus the better performance will be achieved by our system.
However, increasing the sample length will incur more computation complexity and time consumption. 
Thus, a short sample that guarantees high performance is desired.
For this purpose, we first extract fingerprint samples with varying lengths by setting  $M$ to 2, 4, 6, and then increase $M$ by a stride of 6 for values between 6 and 36.
Then, we train and test our system on corresponding samples from dataset A.
As shown in Fig.~\ref{fig:parameter_determination}, our system's accuracy and TNR generally increase with the sample length.
This is because a sample with a larger length can provide more fingerprinting information, thus leading to higher performance.
However, when the sample length exceeds 6, the system performance becomes steady, and a larger length only brings minor accuracy and TNR improvements.
Based on the above results, we set six multimodal data frames in each sample.
In this setting, SecureLink has an overall accuracy of 98.61\% and a TNR of 99.04\% in the closed-world testing.

After determining the sample length $M$, we further report the system performance for each UAV on dataset A.
For each UAV, a OC-SVM is trained using the embeddings of the training and validation data.
Subsequently, the embeddings of the test samples are used for evaluation.
Since our OC-SVMs are designed for open-world authentication, a sample may be classified as a positive by multiple OC-SVMs, but only one classification result is correct.
For this reason, we selected the OC-SVM with the highest matching score as the final prediction result. 
The prediction results of all OC-SVMs are presented in the confusion matrix in Fig.~\ref{fig:closed-world_performance}.
For each OC-SVM, at least 97\% of test samples are correctly classified, and only a small fraction are incorrectly accepted by other OC-SVMs.
In addition, SecureLink obtains over 98.5\% accuracy for authenticating the DJI Phantom 4 Pro drones with the ESP32 SoC and the ESP32-S3 SoC, indicating that they are significantly different from the Tello drones in RF and MEMS imperfections.
The results demonstrate the effectiveness of our system in discriminating between UAVs of different models.

\begin{table}[t]
    \centering
    \caption{System Performance in the Open World}
    \label{tab:atack-experiment}
    \setlength{\tabcolsep}{2mm}{
    \begin{tabular}{ccccc}
    \toprule Round No. & Impersonating UAV & Accuracy & TNR\\ 
    \midrule 
       1& \#0, \#19 & 97.49\% & 96.52\% \\ 
       2& \#6, \#15 & 97.06\% & 96.84\% \\ 
       3& \#13, \#17 & 97.92\% & 97.15\% \\ 
       4& \#7, \#10  & 97.41\% & 97.52\% \\  
       5& \#2, \#5 & 96.97\% & 96.23\% \\ 
       6& \#20, \#21 & 98.42\% & 97.47\% \\  %
       Average &  & \textbf{97.54}\% & \textbf{96.95}\% \\  %
    \bottomrule 
    \end{tabular}
    }
\end{table}

\textbf{Open-World Performance.}
Then, we present the system's performance in the open-world setting, where unseen impersonating UAVs could initiate the authentication process with a fake ID and the stolen certificate.
To fully evaluate our system, we conduct six rounds of open-world tests.
In each round, we take two UAVs as impersonating ones, and their samples are randomly labeled with a registered ID. 
The remaining twenty UAVs are considered legitimate, whose samples are assigned true IDs.
In this way, we obtain multiple versions of dataset B and train and test our system on them.
Table~\ref{tab:atack-experiment} shows the details of six round tests.
Generally speaking, SecureLink suffers a slight performance degradation in the open world, with a TNR decrease of 2.09\%, especially.
This is because the existence of impersonating UAVs results in fewer training samples and more unauthorized samples in the testing phase, rendering our system harder to learn fingerprinting features. 
Moreover, our system achieves the highest performance in the sixth round with an accuracy of 98.42\% and a TNR of 97.47\%.
The reason is that the impersonating UAVs are the DJI Phantom 4 Pro with the ESP32 SoC and the ESP32-S3 SoC in this case, and belong to different types from legitimate UAVs, i.e., twenty Tell drones.
Due to hardware heterogeneity between different UAV models, the RF and MEMS fingerprints of the two impersonating UAVs are more different from those of the legitimate UAVs, making our system easier to recognize spoofed UAVs of other types.
Despite that, SecureLink has good and stable performance in each round and achieves an accuracy of 97.54\% and a TNR of 96.95\% on average for twenty registered UAVs and two impersonating ones.

\begin{table}[t]
    \centering
    \caption{Comparison with Single-Modality Baselines}
    \label{tab:baseline}
    \resizebox{1.0\linewidth}{!}{
    \begin{tabular}{cccc}
    \toprule Methods & 6 Frames & 24 Frames & 36 Frames \\
    \midrule 
        Gyro bias~\cite{gro-bias} & 38.51\% & 38.89\% & 39.31\% \\ 
        Acc errors~\cite{Acc-error} & 68.75\% & 79.28\% & 84.37\% \\ 
        Micro-CSI~\cite{Micro-CSI} & 18.37\% & 26.11\% & 33.92\% \\
        CSI phase errors~\cite{CSI-Phase-error} & 56.74\% & 64.45\% & 70.55\% \\
        \textbf{SecureLink (Ours)} & \textbf{98.61\%} & \textbf{98.96\%} & \textbf{99.26\%} \\
    \bottomrule 
    \end{tabular}}
\end{table}

\begin{table}[t]
  \centering
  \caption{Comparison with Multimodal Baselines}
  \label{tab: multimodal_baselines}
\setlength{\tabcolsep}{6mm}{
\begin{tabular}{ccc}
\toprule Methods                   & Accuracy         & TNR              \\ \midrule 
Aledhari et al.~\cite{9448699}           & 79.24\%          & 81.32\%          \\
Alzahrani et al.~\cite{alzahrani2024enhancing}          & 92.31\%          & 88.57\%          \\
\textbf{SecureLink(Ours)} & \textbf{98.61\%} & \textbf{99.04\%} \\
\bottomrule 
\end{tabular}
}
\end{table}

\textbf{Comparison with Baselines.}
Next, we compare SecureLink with the existing RF and MEMS fingerprinting approaches.
Specifically, we develop two sensor-based approaches using the output bias of gyroscopes~\cite{gro-bias} and errors of accelerometers~\cite{Acc-error}, respectively.
Moreover, two CSI-based baselines using micro CSI features~\cite{Micro-CSI} and phase errors~\cite{CSI-Phase-error} are built.
We evaluate four baselines with different sample lengths on dataset A and report their performance in Table~\ref{tab:baseline}.
As the table shows, as the sample length increases, their performance becomes better.
However, the baseline~\cite{gro-bias} achieves bad performance with an accuracy lower than 40\% in each setting.
This is because this baseline relies on zero-rate sensor outputs and requires the device under test to be stationary, which is not suitable for UAVs with high mobility.
Moreover, Micro-CSI~\cite{Micro-CSI} also shows a low authentication performance, because this approach requires hundreds of CSI measurements to generate a robust fingerprint and cannot extract effective information from limited CSI frames.
Although the remaining two baselines have much better performance than the above two, their performance is not comparable to SecureLink.
This is due to the fact that RF and MEMS fingerprints are complementary, and more robust features can be effectively extracted by our multimodal fusion scheme.
The above results show that SecureLink can achieve better authentication performance with fewer data frames, leading to less waiting time for authenticating each association request.

Moreover, we compare our system with other multimodal fusion methods.
Specifically, the work~\cite{9448699} extracts features from both RF data and image data using two DNNs, respectively, and fuses them using another DNN for UAV detection.
The work~\cite{alzahrani2024enhancing} concatenates data from multiple sensors, such as cameras and GPS sensors, with different weights and feeds them into a SVM for detecting anomaly UAV behaviors, like GPS spoofing and communication jamming.
We build two baselines using their architectures~\cite{9448699,alzahrani2024enhancing} and train and test them on Dataset A.
As Table~\ref{tab: multimodal_baselines} shows, SecureLink outperforms the two baselines, indicating the effectiveness of our multimodal fusion and UAV identification.

\begin{figure}[t]
    \centering
	\includegraphics[width=1.0\linewidth]{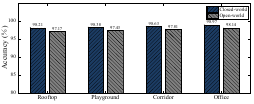}
	\caption{Performance in different environments.}
	\label{fig:enviroments_compare}
\end{figure}

\begin{figure}[t]
    \centering
	\includegraphics[width=1.0\linewidth]{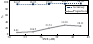}
	\caption{System performance under different SNRs.}
	\label{fig:impact of snrs}
\end{figure}

\textbf{Impact of Surrounding Environments.}
Furthermore, we study the impact of the surrounding environments on our system.
Generally, CSI measurements are sensitive to environmental changes, thus hampering system performance.
To do this, we report the authentication accuracy in four different environments in both the closed-world and open-world settings in Fig.~\ref{fig:enviroments_compare}. 
The results indicate that our system maintains good performance in the two scenarios, and it obtains an accuracy of around 98\% in each environment.
This is attributed to the effective attention-based feature fusion, which extracts complementary information from CSI and telemetry data, thus boosting the system's robustness to environmental changes.

\textbf{Impact of Signal-to-Noise Ratios (SNRs).}
Now, we show system performance under different SNRs on Dataset A.
For each CSI measurement, the SNR can be calculated by subtracting the background noise power from the received signal strength in dBm.
Then, we average all SNRs in one CSI frame and use the mean result to represent the SNR of the corresponding sample.
In this way, we compute the SNRs of all testing samples, group the samples using their SNRs, and evaluate our model in each group.
As depicted in Fig.~\ref{fig:impact of snrs}, the SNRs of testing samples fall between 30dB and 45dB, and more than 50\% of them have SNRs greater than 39dB.
As the SNR increases from 30dB to 45dB, the authentication accuracy improves slowly from 96.37\% to 98.76\%.
The marginal performance improvement suggests the high robustness under dynamic wireless channels.

\begin{figure}[t]
  \centering
  \includegraphics[width=\linewidth]{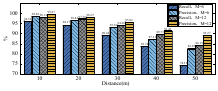} 
  \caption{Performance in different communication distances}
  \label{fig:dist_extra} 
\end{figure}

\textbf{Impact of Communication Distance.}
We investigate the impact of the UAV-GCS communication distance on SecureLink.
To do this, twenty DJI Tello drones are placed 10m, 20m, 30m, 40m, and 50m away from the GCS.
For each drone, we control it under different motion states and collect CSI and MEMS data, lasting about 500 seconds at each distance.
Using the collected data, we conduct two sets of experiments to investigate the communication distance thresholds that can guarantee performance requirements. 
In these two sets of experiments, our system is trained and tested based on two sample lengths, i.e., 6 and 12.
The system's recall and precision are reported in Fig.~\ref{fig:dist_extra}.
The system performance degrades as the distance increases.
This is because an increase in communication distance decreases the SNR, thereby worsening the quality of RF fingerprints in the received signals.
When $M=6$, the precision of our SecureLink remains high (above 93\%) if a communication distance is less than 30m. 
Beyond this 30m threshold, the precision decreases significantly. 
When $M=12$, the precision remains above 91\% if less than 40m. 
Moreover, a clear improvement can be achieved by increasing the sample length from 6 to 12. 
Especially at 50m, a precision increase of about 7\% is achieved.
The above observations suggest a trade-off between the communication distance and the sample length.

\begin{figure}[t]
    \centering
	\includegraphics[width=1.0\linewidth]{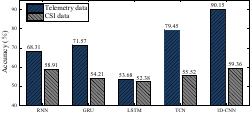}
	\caption{Performance of five neural networks on CSI and telemetry data.}
	\label{fig:networks_compare}
\end{figure}

\textbf{Impact of 1D-CNN.}
Next, we verify the unimodal feature extraction capability of the adopted 1D-CNN network.
For this purpose, we choose four other networks commonly used for feature extraction of time-series data.
They are a recurrent neural network (RNN), a gate recurrent unit (GRU), an LSTM, and a temporal convolutional network (TCN).
The 1D-CNN and TCN are configured with two convolutional layers, while the RNN, GRU, and LSTM have two recurrent layers.
All networks are configured with a metric embedding layer each.
We train and test them on the two modalities of samples from dataset A.
As shown in Fig.~\ref{fig:networks_compare}, most networks have an accuracy between 50\% to 60\% on the CSI data.
However, LSTM has the lowest accuracy, with 53.68\% on telemetry data and 52.38\% on CSI data. 
In addition, it can be observed that 1D-CNN achieves the highest accuracy on the CSI and telemetry data, which demonstrates its high effectiveness in unimodal feature extraction.

\begin{figure}
    \centering
	\includegraphics[width=1.0\linewidth]{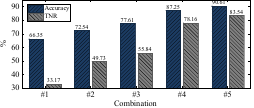}
	\caption{Performance of 1D-CNN on five combinations of MEMS features.}
	\label{fig:sensors_compare}
\end{figure}

\textbf{Impact of Telemetry Data.}
Then, we validate the impact of the selected eight sensory fields from telemetry data.
Specifically, these MEMS sensor data can be categorized into three types.
The first type is the attitude of UAVs, including pitch, roll, and yaw.
The second type is the altitude, consisting of outputs of the ToF sensor and barometer.
The third type is the three-axis accelerometer.
Thus, we make five combinations: 1) Att \& Acc; 2) ToF \& Baro; 3) Att \& ToF \& Baro; 4) Acc \& ToF \& Baro; 5) and All.
Based on dataset A, we extract such features from samples with a length of six data frames.
Then, 1D-CNN is exploited to perform authentication using different features because it achieves the best performance in unimodal feature extraction, as mentioned above.
The result is depicted in Fig.~\ref{fig:sensors_compare}.
Generally, the authentication performance boosts as the number of sensory fields increases. 
The accuracy of barometer and ToF data, achieving 72.54\%, appears to be more effective than that of accelerometer and attitude data, which reaches to 66.35\%.
When it comes to the combinations of three types, the accuracy of the 1D-CNN is more than 77.61\%.
Additionally, the exploitation of all sensory fields works best and increases the accuracy up to 90.61\% and the TNR up to 83.54\%. 
The above results manifest that all selected sensory fields are useful in characterizing airborne MEMS fingerprints.

\begin{table}[t]
    \centering
    \caption{Ablation Study of Multimodal Fusion}
    \label{tab:fusion-approaches}
    \setlength{\tabcolsep}{2mm}{
    \begin{tabular}{ccc}
    \toprule Architectures & Accuracy & TNR \\
    \midrule
        OC-SVM & 53.25\% & 8.61\% \\ 
        1D-CNN + OC-SVM & 96.46\% & 95.93\% \\ 
        1D-CNN + BiLSTM + OC-SVM & 97.64\% & 97.86\% \\ 
        1D-CNN + BiLSTM + Attention + OC-SVM & 98.61\% & 99.04\% \\
    \bottomrule 
    \end{tabular}}
\end{table}

\textbf{Ablation Study.}
We show the feature extraction ability of each component in multimodal fusion.
We compare three architectures: 1D-CNN, 1D-CNN$+$BiLSTM, and 1D-CNN$+$BiLSTM$+$Attention.
It is worth noting that, different from the above 1D-CNN that uses either CSI or telemetry data alone, the 1D-CNN here refers to two 1D-CNN branches that use the two modalities as input, respectively.
Moreover, each architecture is followed by OC-SVM for effective classification.
We train and test three architectures and a OC-SVM alone on dataset A.
The experimental results are reported in Table~\ref{tab:fusion-approaches}.
For raw CSI and telemetry features, the single OC-SVM struggles to recognize negative samples with a low TNR of 8.61\%, indicating that proper feature representation is necessary.
Moreover, our 1D-CNN brings a huge performance improvement of more than 40\% in terms of accuracy, suggesting its ability to extract unimodal features.
In addition, our BiLSTM and multi-head attention can further discover underlying identity information and boost authentication performance.

\begin{figure}[t]
    \centering
    \subfigure[Histogram of phase difference variances.]{
    	\includegraphics[width=0.46\columnwidth]{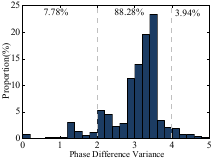}
    	\label{fig:tof_outlier}
     }
     \subfigure[Performance under different scaling parameters.]{
     \includegraphics[width=0.46\columnwidth]{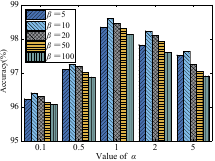}
     \label{fig:aligned_csi}
     }
    \caption{Results of parameter determination.} 
    \label{fig:hyper_parameter_determination}
\end{figure}

\textbf{Parameter Determination.}
We now explain how to determine the threshold $\eta$ in RF fingerprint extraction and the scaling parameters $\alpha$ and $\beta$ in UAV registration.
To determine the value of $\eta$, we compute phase variances for all collected CSI measurements and plot their histogram in Fig.~\ref{fig:hyper_parameter_determination}~(a).
As the figure shows, more than 80\% of phase variances fall between 2 and 4, and more than 95\% of phase variances are smaller than 4.
Because the outliers caused by environmental interference usually have steep variances, we hence set $\eta = 4 $ to retain the majority of CSI measurements and filter out those with high variances.
To obtain the values for $\alpha$ and $\beta$, we vary $\alpha$ over the set $\left\{0.1, 0.5, 1, 2, 5\right\}$ and $\beta$ over $\left\{5, 10, 20, 50, 100\right\}$ simultaneously and train our model in each case.
As depicted in Fig.~\ref{fig:hyper_parameter_determination}~(b), as $\alpha$ increases, the system performance improves initially.
But when it is greater than 1, the system performance degrades slowly.
Similar trends can be observed when $\beta \neq 10 $ in each case.
Based on the above results, we set $\alpha = 1$ and $\beta = 10$ in our experiments.

\begin{table}[t]
    \centering
    \caption{Test Runtime of SecureLink}
    \label{tab:time-cost}
    \setlength{\tabcolsep}{2mm}{
\begin{tabular}{ccccc}
\hline
Platform      & Preprocessing &Fusion& Identification & Total \\ \hline
Legion Y9000X &       13.76ms        &                        0.33ms                                     &                     0.54ms                                           &    14.63ms   \\
Youyeetoo X1     &     15.64ms          &                 0.47ms                                             &                      0.61ms                                          &   16.72ms   \\ \hline
\end{tabular}
}
\end{table}

\textbf{Time Consumption.}
Finally, to verify the real-time ability of our system, we demonstrate its runtime on a Lenovo Legion Y9000X laptop and a Youyeetoo X1 development board.
After implementing our system on the two platforms, we generate 2K samples from the collected multimodal data, feed them into our system, and record the average runtime of three components, i.e., data preprocessing, multimodal fusion, and UAV identification, for each test sample.
As reported in Table~\ref{tab:time-cost}, the laptop is faster than the development board in each process, because it has better computation ability.
Moreover, most of the time consumption is attributed to data preprocessing, about 13.76ms on Legion Y9000X and 15.64ms on Youyeetoo X1, and the total time for verifying a sample is around 15ms.
The above results suggest that our system has a quick response to an authentication request.
Note that the time cost of data collection is about 600ms because the Tello drone sends telemetry readings at a frequency of 10~Hz, and our system relies on six frames for authentication.

\section{Discussion}

\textbf{Robustness to UAV Model Variations.}
SecureLink relies on RF and MEMS data, which are fundamental in UAV systems. 
In terms of RF signals, although our study in the paper is based on the 802.11/WiFi standard, SecureLink can also be applied to UAVs using other protocols, such as Bluetooth, ZigBee, and LoRa.
This is because these protocols can also provide channel state information to SecureLink.
As for MEMS data, SecureLink can use readings from accelerometers, barometers, and ToF sensors as MEMS data. 
These meters and sensors are basic components in flight control systems.
If some sensors are missing in some models, the dimension of the MEMS data will decrease.
While such changes may affect the number of data frames required, they will not alter the UAV identifications or the subsequent workflow of multimodal fusion. 
This suggests that SecureLink has high robustness to variations in UAV models.

\textbf{Compatibility with Different Drone Vendors.}
Many aviation authorities around the world are promoting the use of standardized communication protocols (e.g., MAVLink) to ensure the safety and interoperability of UAV operations.
For example, the FAA's Remote ID rule mandates that drones must broadcast identification and flying status information~\cite{FAA}.
SecureLink aligns well with these regulatory trends and takes advantage of the already available data required by such standardized protocols. 
Thus, our system is compatible with drone vendors that adopt standardized UAV communication protocols.

\section{Related Work}\label{sec:related work}

\textbf{UAV Detection and Classification.}
UAV detection and classification techniques are mainly based on radar Doppler, RF, acoustic, and optical signatures.
The work~\cite{radar1} achieves the distinction between UAVs and other objects by analyzing micro-Doppler features.
The work~\cite{r10} analyzes radar signals by wavelet decomposition for UAV classification.
Radar-based technologies are susceptible to the target radar cross-section and other clutter.
Moreover, radar equipment is generally bulky and hard to deploy and maintain.
Recently, the exploitation of RF features for UAV detection and classification has become increasingly popular~\cite{r6,r11}.
The authors~\cite{rf1} detect and classify UAVs by extracting physical characteristics of UAVs' body vibration and body shifting from signals transmitted by UAVs.
The work~\cite{csi_det} proposes a CSI-based approach for UAV detection. 
Acoustic signatures from Mel-frequency cepstral coefficients are used to detect the UAV presence and identify the model of UAVs~\cite{r8}.
UAV classification using optical features is mainly achieved by neural networks~\cite{opromolla2018vision,vison1,vison2,vison3}.
These networks can autonomously learn the relevant features from the captured UAV images or videos.
However, the above approaches focus on detecting or classifying UAVs and are not suited to securing UAV communications.

\textbf{UAV Authentication.}
Existing UAV authentication methods mainly adopt acoustic characteristics, RF characteristics, and MEMS characteristics.
The authors~\cite{acoustic2} compute Mel-frequency cepstral coefficients (MFCC), delta-MFCC, and delta-delta MFCC from the combined motor and propeller noise of each UAV as acoustic fingerprints, and apply quadratic discriminant analysis to authenticate UAVs.
However, this method requires the device under test to be in proximity and is prone to environmental noise.
In recent years, some researchers~\cite{r5,9010185,droneauth1, r7} have leveraged RF fingerprints in fine-grained IQ samples for UAV recognition.
Fractal dimension, axially integrated bispectra, and square integrated bispectra are extracted from time-domain RF signals as UAV fingerprints, and an extreme learning machine classifier is used for UAV identification~\cite{r5}.
The work extracts cyclostationary spectral-correlation features and statistical RF signatures and employs an SVM to identify UAVs~\cite{9010185}.
Yazdinejad et. al.~\cite{droneauth1} extract various orthogonal frequency division multiplexing-based RF features and leverage a federated learning framework for privacy-preserving drone authentication.
The approach feeds raw IQ samples into a multi-channel neural network with data augmentation, then applies a two-step score-based aggregation method to fingerprint different UAV models~\cite{r7}.
However, IQ samples require dedicated devices, such as universal software radio peripherals.
In contrast, we leverage easily obtainable CSI measurements~\cite{29,30,picoscenes} for UAV authentication.
The work~\cite{9951057} exploits raw MEMS sensor readings as UAV fingerprints, then designs a challenge-response authentication protocol to identify authorized UAVs.
Moreover, the work~\cite{gro-bias} extracts MEMS imperfections from stand-alone gyroscope sensors for UAV authentication.
However, the above approach focuses on the stationary scenario and may not generalize well to various flying states.

\section{Conclusion}\label{sec:conclusion}
This paper presents SecureLink, a novel cross-layer UAV authentication system that exploits RF and MEMS fingerprints for securing UAV communications.
We found that freely available CSI and telemetry measurements contain useful information regarding airborne hardware imperfections that can be fused to efficiently and accurately authenticate commercial-off-the-shelf UAVs in the real world.
SecureLink first aligns fingerprints from CSI and telemetry measurements. 
Then, an attention-based neural network is devised for in-depth feature fusion. 
With these developments, the fused features are fed into a OC-SVM classifier for open-world authentication.
The whole process only requires six additional data frames for feature fusion and UAV authentication.

We implement SecureLink on three different types of UAVs, including twenty DJI Tello drones, one DJI Phantom 4 Pro drone with the ESP32 SoC, and one DJI Phantom 4 Pro drone with the ESP32 S3 SoC, respectively. 
We collect multimodal data, lasting a total of about twelve hours, in real-world environments.
The evaluation results demonstrate that SecureLink achieves a closed-world accuracy of 98.61\% and an open-world accuracy of 97.54\%, outperforming the existing approaches in robustness and communication overheads.
We believe that SecureLink can be employed beyond UAV communication systems, extending its applicability to other domains, such as vehicle communication systems, for the identification of impersonating devices.

\bibliographystyle{ieeetr}
\bibliography{IEEEabrv,./reference}

\end{document}